\documentclass[12pt]{article}
\usepackage{amsmath}
\usepackage{amssymb}
\usepackage[dvips]{graphicx}
\usepackage{amsmath,amscd}
\usepackage{color}
\usepackage{mathrsfs}
\usepackage{latexsym}
\makeatletter

  \@addtoreset{equation}{section}
\makeatother
\allowdisplaybreaks[4]

\begin{document}
\begin{titlepage}
\thispagestyle{empty}
\begin{flushright}
NCTS-TH/1502

\end{flushright}
\vspace{3mm}

\begin{center}
{\LARGE Inflation from radion gauge-Higgs potential\vspace{0.1cm} \\
at Planck scale}

\end{center}

\begin{center}
\lineskip .45em
\vskip1.5cm
{\large Yugo Abe$^a$\footnote{E-mail: 13st301k@shinshu-u.ac.jp }, 
Takeo Inami$^{b}$\footnote{E-mail: inami@phys.chuo-u.ac.jp}, 
Yoshiharu Kawamura$^a$\footnote{E-mail: haru@azusa.shinshu-u.ac.jp}
and Yoji Koyama$^{c}$\footnote{E-mail: ykoyama@phys.nthu.edu.tw}}

\vskip 1.5em
${}^a\,${\large\itshape Department of Physics, Shinshu University, Matsumoto 390-8621, Japan}\\[1mm]
${}^b\,${\large\itshape Mathematical Physics Lab., Riken Nishina Center, Saitama, Japan}\\[1mm]
${}^b\,${\large\itshape Department of Physics, National
Taiwan University, Taipei, Taiwan}\\[1mm]
${}^c\,${\large\itshape National Center for Theoretical Science, Hsinchu, Taiwan 300}

 \vskip 4.5em
\end{center}
\begin{abstract}

We study whether the inflation is realized based on the radion gauge-Higgs potential obtained from the one-loop calculation in the 5-dimensional gravity coupled to a $U(1)$ gauge theory.
We show that the gauge-Higgs can give rise to inflation in accord with the astrophysical data and the radion plays a role in fixing the values of physical parameters.
We clarify the reason why the radion dominated inflation and the hybrid inflation cannot occur in our framework.

\end{abstract}
\end{titlepage}

\newpage

\abovedisplayskip=1.0em
\belowdisplayskip=1.0em
\abovedisplayshortskip=0.5em
\belowdisplayshortskip=0.5em
\section{\it Introduction}

It is believed that the universe has evolved into the present one 
as a result of a rapid expansion called $\lq\lq$inflation'' in its very early stage.
The well-known difficulties in the standard big-bang model are 
solved by the slow-roll inflation scenario~\cite{LR,HKO}.
Furthermore recent accurate measurements have confirmed 
the predictions of a flat universe 
with a nearly scale-invariant density perturbation~\cite{Planck2015}.

A remaining big problem is to explore the origin of inflation or
to disclose the identity of a scalar particle called $\lq\lq$inflaton''.
The following requirements can be imposed on the model as clues to a solution.
The inflaton should be the inevitable product of a theory at a high-energy scale.
The potential of inflaton should be stable against various corrections.
Concretely, there should be no fine tuning among parameters
receiving quantum corrections and no sensitivity to gravitational corrections. 

On a 4-dimensional (4D) space-time, 
no model has yet been known to solve the problem completely.
In most models, the inflaton is an ad-hoc particle introduced by hand,
and the stability of potential is threatened
by radiative corrections and gravitational ones relating non-renormalizable terms
suppressed by the power of the Planck mass
because the inflaton takes a larger value than the Planck mass
and such corrections cannot be controlled
without any powerful symmetries and/or mechanisms.

Effective field theories on a higher-dimensional space-time 
provide a possible solution to the problem.
Some scalar fields exist inevitably as parts of ingredients in the theory 
and are massless at the tree level.
The scalar potential can be induced radiatively
and stabilized by local symmetries.
Typical ones are the extranatural inflation model~\cite{ACCR,IKLM}
and the radion inflation model~\cite{FIK}.
In the former model,
a scalar field called $\lq\lq$gauge-Higgs'' appears from
the extra-dimensional component(s) of the gauge field.
It becomes dynamical degrees of freedom called the Wilson line phase 
and its value is fixed by quantum corrections~\cite{Hosotani}.
In the latter model,
a scalar field called $\lq\lq$radion'' originates from 
the extra-dimensional component(s) of the graviton,
and its vacuum expectation value (VEV) is related to the size of the extra space.

Recently, the effective potential with respect to both the radion 
and the gauge-Higgs has been derived at the one-loop level
upon the $S^1$ compactification,
from the gravity theory coupled to a $U(1)$ gauge boson and matter fermions 
on a 5-dimensional (5D) space-time~\cite{AIKK}.
We refer to the potential as the $\lq\lq$radion gauge-Higgs potential''.
It is interesting to investigate whether it works as the inflaton potential
and what features exist in such a coexisting system.

In this article, we study whether the slow-roll inflation is realized
compatible with the astrophysical data,
based on the radion gauge-Higgs potential.
In the analysis, we pay attention to
which particle can play the role of inflaton
and the magnitude of physical parameters.

The content of our article is as follows.
In the next section, we explain the radion gauge-Higgs potential 
and constraints on the inflaton potential.
In Sect. 3, we study the gauge-Higgs inflation
after focusing on a candidate of inflaton.
It will be shown that the inflation can be achieved in our framework.
In the last section, we give conclusions and discussions.

\section{\it Setup}
\subsection{\it Radion gauge-Higgs potential}

Based on a 5D theory containing a graviton, a $U(1)$ gauge boson and fermions,
the following radion gauge-Higgs potential have been obtained 
at the one-loop level~\cite{AIKK},
\begin{eqnarray}
&&V(\chi,\varphi)=\frac{3}{\pi^{2}\chi^{2}L^{4}}\left[-2\zeta(5)
+c_{1}\sum^{\infty}_{k=1}\left(\frac{1}{k^{5}}+r_{m}\frac{Lm\chi^{1/3}}{k^{4}}+r^{2}_{m}\frac{L^{2}m^{2}\chi^{2/3}}{3k^{3}}\right)e^{-kr_{m}Lm\chi^{1/3}}~~~~~~~\right.\nonumber\\
&&~~~~~\left.+c_{2}\sum^{\infty}_{k=1}\left(\frac{1}{k^{5}}+\frac{Lm\chi^{1/3}}{k^{4}}+\frac{L^{2}m^{2}\chi^{2/3}}{3k^{3}}\right)e^{-kLm\chi^{1/3}}\cos{(kg_{4}L\varphi)}\right]+aL\chi^{-1/3}+ \cdots~,
\label{Veff}
\end{eqnarray}
where $\chi$ is the radion, $\varphi$ the gauge-Higgs,
$L$ the $S^{1}$ circumference,
$g_{4}$ the 4D gauge coupling constant,
$c_{1}$ and $c_{2}$ the numbers of neutral and charged fermions
whose masses are $\mu$ and $m$, $r_{m}=\mu/m$
and the ellipsis stands for terms including infinities 
whose form is consistent with the general covariance, e.g. $\Lambda^5 L\chi^{-1/3}$ with $\Lambda$ the UV cutoff.
Using $r_m$, we can control contributions of two kinds of fermions 
to the potential through the factor $e^{-kr_{m}Lm\chi^{1/3}}$.
Note that the 5D cosmological constant counter term
$aL\chi^{-1/3}$ is introduced to remove infinities~\cite{A&C, P&P}.

The stabilization of $\chi$ and $\varphi$ has been studied 
and it has been shown that the potential has a stable minimum if $c_{1}>c_{2}+2$.
The physical length of the 5-th dimension $L_{\rm phys}$,
the physical fermion masses $m_{\rm phys}$ and $\mu_{\rm phys}$ are given by
\begin{eqnarray}
L_{\rm phys}=L\langle\chi\rangle^{1/3}~,
~~m_{\rm phys}=m\langle\chi\rangle^{-1/6}~,
~~\mu_{\rm phys}=r_{m}m_{\rm phys}~,
\label{physical para}
\end{eqnarray}
where $\langle\chi\rangle$ is the VEV of $\chi$ at the minimum of $V$.

It is an attractive idea that
the structure and evolution of our 4D space-time
might be deeply linked to the dynamics of quantities relating an extra space. 
The $V(\chi,\varphi)$ must describe the physics around the Planck scale
if the magnitude of $L_{\rm phys}^{-1}$, $m_{\rm phys}$ and/or $\mu_{\rm phys}$
can be comparable to the reduced Planck mass 
$M_{G}$ ($\equiv(8\pi G)^{-1/2} = 2.4 \times 10^{18}$GeV).
There are nearly flat domains around the minimum of $V(\chi,\varphi)$.
Hence it is reasonable to expect that the radion gauge-Higgs potential 
plays the role of inflaton potential.

\subsection{\it Constraints on inflaton potential}

From the inflation theory and the observational data, we have several constraints 
on the inflaton potential $V$~\cite{LR,Planck2015}.
Leading constraints are listed as
\begin{eqnarray}
&\cdot&{\rm the~slow\mathchar`-roll~parameters}~~~~~~~~\epsilon\equiv\frac{1}{2}M^{2}_{G}\left(\frac{V'}{V}\right)^{2}\ll1~,~~~\eta~\equiv~M^{2}_{G}\frac{V''}{V}\ll1~,
\label{slow-roll para}\\
&\cdot&{\rm the~spectral~index}~~~~~~~~~~~~~~~~n_{s\ast}\equiv
1-6\epsilon_{\ast}+2\eta_{\ast}=0.9655\pm0.0062~,\label{spectral index}\\
&\cdot&{\rm the~number~of~e\mathchar`-foldings}~~~~~~~N\equiv\frac{1}{M^{2}_{G}}\left|\int^{\phi_{e}}_{\phi_{\ast}}\frac{V}{V'}d\phi \right|=50\sim60~,\label{e-folding}\\
&\cdot&{\rm the~scalar~power~spectrum}~~~~~{\mathcal P}_{\zeta\ast}\equiv
\left. \frac{1}{12\pi^{2}M^{6}_{G}}\frac{V^{3}}{(V')^{2}}\right|_{\phi = \phi_{\ast}}
\!\!\!\!\!=(2.196\pm0.079)\times10^{-9}\label{Power spectrum},\\
&\cdot&{\rm the~tensor\mathchar`-to\mathchar`-scalar~ratio}~~~~~~~
r = 16 \epsilon_{\ast} < 0.12~,\label{tensor}\\
&\cdot&{\rm the~vacuum~energy}~~~~~~~~~~~~~~~V(\phi_e) \cong 0~,\label{Vacuum energy}
\end{eqnarray}
where $\phi$ is the inflaton field canonically normalized with a mass dimension, $V'=\partial V/\partial\phi$ and $V''=\partial^{2}V/\partial\phi^{2}$.
The quantities at the horizon exit are indicated by $\ast$.
The subscript $e$ denotes the value at the end of the slow-roll 
when $\epsilon\sim1$ or $\eta\sim1$.

Our 4D effective theory contains two scalar fields inevitably.
One is the canonically normalized radion $\chi'$ defined by
\begin{eqnarray}
\chi'\equiv\frac{M_{G}}{\sqrt{6}}\ln{\chi}~.
\label{canonicalscalar}
\end{eqnarray}
The other is the gauge-Higgs $\varphi$ given as a canonically normalized one.
In the analysis of the slow-roll inflation,
it is often convenient to use the dimensionless field variables
such as $x (=L^{3}m^{3}\chi)$ and the Wilson line phase $\theta (=g_{4}L\varphi)$.
Using them, the potential (\ref{Veff}) is rewritten as
\begin{eqnarray}
V(x,\theta)&=&\frac{3L^{2}m^{6}}{\pi^{2}x^{2}}\left[-2\zeta(5)+c_{1}\sum^{\infty}_{k=1}\left(\frac{1}{k^{5}}+r_{m}\frac{x^{1/3}}{k^{4}}+r^{2}_{m}\frac{x^{2/3}}{3k^{3}}\right)e^{-kr_{m}x^{1/3}}\right.\nonumber\\
&~&~~~\left.+c_{2}\sum^{\infty}_{k=1}\left(\frac{1}{k^{5}}+\frac{x^{1/3}}{k^{4}}+\frac{x^{2/3}}{3k^{3}}\right)e^{-kx^{1/3}}\cos{(k\theta)}\right]+\frac{L^{2}m}{x^{1/3}}a+ \cdots~.
\label{V}
\end{eqnarray}

The constraint (\ref{Vacuum energy}) means 
that the inflation ends at $\phi = \phi_e$,
and it is given by the renormalization condition $V(\langle\chi\rangle,\langle\varphi\rangle)=0$
because the inflation must be terminated near the minimum of $V$.
Here, $\langle\varphi\rangle$ is the VEV of $\varphi$ at the minimum of $V$.

\section{\it Analysis}
\subsection{\it Candidate of inflaton}

Let us guess the inflaton $\phi$
under the assumption that $V(x, \theta)$ is the inflaton potential.
There are several possibilities.
(a) $\phi$ is a mixture of $x$ and $\theta$.
(b) $\phi$ is $x$ or $\theta$.
In the hybrid inflation scenario~\cite{Linde:hybrid},
there is a field called $\lq\lq$waterfall'' $\omega$ other than $\phi$. 
The role of $\omega$ is to terminate the inflation induced by $\phi$,
through the fall into the minimum of $V$.

The candidate of $\phi$ is focused on by the slow-roll conditions (\ref{slow-roll para}).
The first and second derivatives of $V$ with respect to $\chi'$ and $\varphi$
are given by
\begin{eqnarray}
&~&\frac{\partial V}{\partial\chi'}
=\frac{\sqrt{6}x}{M_{G}}\frac{\partial}{\partial x}V(x, \theta)
=-\frac{6\sqrt{6}L^{2}m^{6}}{M_{G}\pi^{2}x^{2}}\left[\frac{\pi^{2}x^{5/3}}{18m^{5}}a-2\zeta(5)\right.
\nonumber \\
&~& ~~~~~~~~~~~~~~ + c_1 \sum^{\infty}_{k=1}\left(\frac{1}{k^{5}}+r_{m}\frac{x^{1/3}}{k^{4}}
 +r^{2}_{m}\frac{7x^{2/3}}{18k^{3}}+r^{3}_{m}\frac{x}{18k^{2}}\right)
e^{-kr_{m}x^{1/3}} \nonumber\\
&~&~~~~~~~~~~~~~~ \left.+ c_2 \sum^{\infty}_{k=1}\left(\frac{1}{k^{5}}+\frac{x^{1/3}}{k^{4}}+\frac{7x^{2/3}}{18k^{3}}+\frac{x}{18k^{2}}\right)e^{-kx^{1/3}}\cos{(k\theta)}\right],
\label{deriv3}\\
&~&\frac{\partial^{2} V}{\partial\chi'^{2}}
=\frac{6x}{M_{G}^{2}}\frac{\partial}{\partial x}V(x, \theta)+\frac{6x^{2}}{M_{G}^{2}}\frac{\partial^{2}}{\partial x^{2}}V(x, \theta)
=\frac{72L^{2}m^{6}}{M_{G}^{2}\pi^{2}x^{2}}
\left[\frac{\pi^{2}x^{5/3}}{108m^{5}}a-2\zeta(5)\right.
\nonumber \\
&~& ~~~~~~~~~~~~~~ + c_1 \sum^{\infty}_{k=1}
\left(\frac{1}{k^{5}}+r_{m}\frac{x^{1/3}}{k^{4}}+r_{m}^{2}\frac{23x^{2/3}}{54k^{3}}+r_{m}^{3}\frac{5x}{54k^{2}}+r_{m}^{4}\frac{x^{4/3}}{108k}\right)e^{-kr_{m}x^{1/3}}
\nonumber\\
 &~&~~~~~~~~~~~~~~ \left.+ c_2 \sum^{\infty}_{k=1}\left(\frac{1}{k^{5}}+\frac{x^{1/3}}{k^{4}}+\frac{23x^{2/3}}{54k^{3}}+\frac{5x}{54k^{2}}+\frac{x^{4/3}}{108k}\right)
e^{-kx^{1/3}}\cos{(k\theta)}\right]~,
\label{deriv4}\\
&~&\frac{\partial V}{\partial\varphi}
=\frac{1}{f}\frac{\partial}{\partial\theta}V(x, \theta)
=-\frac{3L^{2}m^{6}}{f\pi^{2}x^{2}} 
c_2 \sum^{\infty}_{k=1}\left(\frac{1}{k^{4}}+\frac{x^{1/3}}{k^{3}}
+\frac{x^{2/3}}{3k^{2}}\right)e^{-kx^{1/3}}\sin{(k\theta)}~,
\label{deriv1}\\
&~&\frac{\partial^{2} V}{\partial\varphi^{2}}
=\frac{1}{f^{2}}\frac{\partial^{2}}{\partial\theta^{2}}V(x, \theta)
=-\frac{3L^{2}m^{6}}{f^{2}\pi^{2}x^{2}} c_2 \sum^{\infty}_{k=1}\left(\frac{1}{k^{3}}+\frac{x^{1/3}}{k^{2}}+\frac{x^{2/3}}{3k}\right)e^{-kx^{1/3}}\cos{(k\theta)}~,
\label{deriv2}\\
&~&\frac{\partial^{2} V}{\partial\chi' \partial\varphi}
=\frac{\sqrt{6}x}{f M_{G}}\frac{\partial^2}{\partial x \partial \theta}V(x, \theta)
\nonumber\\
 &~&~~~~~~~~~
=-\frac{6\sqrt{6}L^{2}m^{6}}{f M_{G}\pi^{2} x^{2}}
c_2 \sum^{\infty}_{k=1} \left(\frac{1}{k^{4}}
+\frac{x^{1/3}}{k^{3}} +\frac{7x^{2/3}}{18k^{2}}+\frac{x}{18k}\right)
e^{-kx^{1/3}} \sin{(k\theta)}~,
\label{deriv5}
\end{eqnarray}
where $f=1/(g_{4}L)$.

We consider the case that both $\chi'$ and $\varphi$ are apart from 
$\langle \chi' \rangle$ and $\langle \varphi \rangle$
and they are traveling toward the minimum of $V$.
Then, using (\ref{deriv3}) -- (\ref{deriv2}),
the counterparts of $\epsilon$ and $\eta$ are estimated as
\begin{eqnarray}
\hspace{-0.8cm}&~& \epsilon_{\chi'} 
\equiv \frac{1}{2}M^{2}_{G}\left(\frac{\partial V/\partial\chi'}{V}\right)^{2}
= O(1)~,~~
\eta_{\chi'} \equiv M^{2}_{G}\frac{\partial^2 V/\partial\chi'^2}{V}
= O(1)~, 
\label{slow-roll-chi}\\
\hspace{-0.8cm}&~& \epsilon_{\varphi} 
\equiv \frac{1}{2}M^{2}_{G}\left(\frac{\partial V/\partial\varphi}{V}\right)^{2}
= O((M_G/f)^2)~,~~
\eta_{\varphi} \equiv M^{2}_{G}\frac{\partial^2 V/\partial\varphi^2}{V}
= O((M_G/f)^2)~, 
\label{slow-roll-varphi}\\
\hspace{-0.8cm}&~& \epsilon_{\chi'\varphi} 
\equiv \frac{1}{2}M^{2}_{G}
\left(\frac{\partial V/\partial\chi'}{V}\frac{\partial V/\partial\varphi}{V}\right)
= O(M_G/f)~,~~
\eta_{\chi'\varphi} \equiv M^{2}_{G}\frac{\partial^2 V/\partial\chi'\partial\varphi}{V}
= O(M_G/f)~. 
\label{slow-roll-chivarphi}
\end{eqnarray}

The $\epsilon$ ($\eta$) can be given by a linear combination of
$\epsilon_{\chi'}$, $\epsilon_{\varphi}$ and $\epsilon_{\chi'\varphi}$
($\eta_{\chi'}$, $\eta_{\varphi}$ and $\eta_{\chi'\varphi}$)
locally on the space of scalar fields.
From (\ref{slow-roll-chi}) -- (\ref{slow-roll-chivarphi}),
we see that the slow-roll conditions are fulfilled
if and only if $\varphi$ dominantly contributes to $\epsilon$ and $\eta$, i.e.,
$\epsilon \simeq \epsilon_{\varphi}$ and $\eta \simeq \eta_{\varphi}$,
and $f$ is much bigger than $M_G$.
In this way, the gauge-Higgs is a candidate of inflaton, while the radion fails to be.

In the case with $f \gg M_G$, the value of $V$ changes rapidly in the direction of $\chi'$
because $|\partial V/\partial \chi'| \gg |\partial V/\partial \varphi|$.
Then, in the first stage, $\chi'$ is moving toward 
the region where $\partial V/\partial \chi' \simeq 0$,
but $\chi'$ cannot play the role of inflaton, in this period,
because $\epsilon_{\chi'} = O(1)$ and $\eta_{\chi'} = O(1)$
except for a narrow region near $\langle \chi' \rangle$.
Hence the hybrid inflation cannot occur in our model.

After $\chi'$ approaches the region where $\partial V/\partial \chi' \simeq 0$ sufficiently,
$\varphi$ also starts to move toward the minimum of $V$ generating inflation.
Then, the inflaton is, in general, a mixture of $\chi'$ and $\varphi$
that follows a path designated by
$\chi' = \chi'(\varphi)$ satisfying $\partial V/\partial \chi' \simeq 0$.
This corresponds to the multi-field inflation.
In this article, we focus on the case that the inflaton is rolling down 
in the almost $\varphi$-direction, corresponding to the single-field inflation, for simplicity.
Concretely, for the trajectory $x = x(\theta$)
derived from  $\partial V/\partial x = 0$,
we study the parameter region that satisfies the condition,
\begin{eqnarray}
r_{V} \equiv \left| \frac{V(x(\theta + \varDelta \theta), \theta) - V(x(\theta), \theta)}
{V(x, \theta + \varDelta \theta) - V(x, \theta)}\right| \ll 1~,
\label{DeltaV}
\end{eqnarray}
where we use $x$ and $\theta$ in place of $\chi'$ and $\varphi$.
We refer to the inflation that the inflaton consists of almost $\varphi$
as the $\lq\lq$gauge-Higgs dominated inflation''.

\begin{figure}[htbp]
\begin{minipage}{0.5\hsize} 
\begin{center}
\includegraphics[width=85mm]{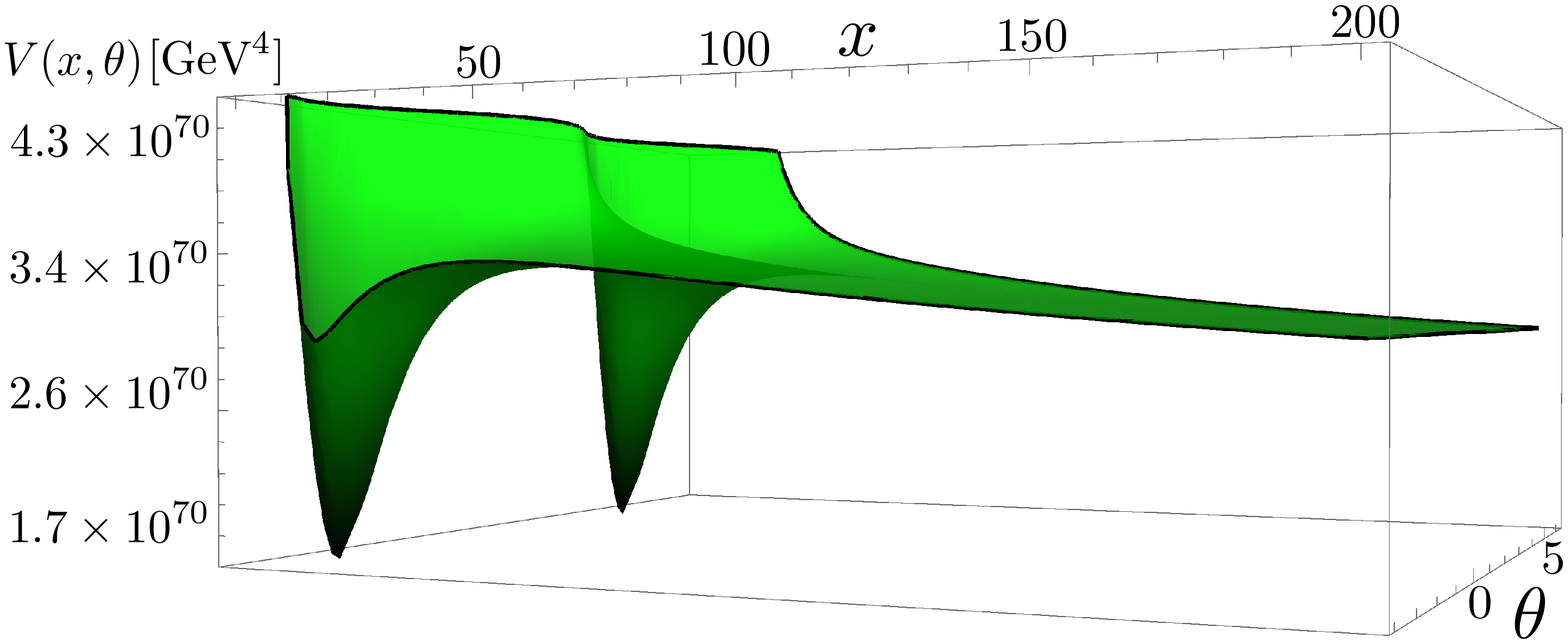}
\end{center}
\vspace{0mm}
\vskip-\lastskip
\caption{Potential for $r_m=1.0$}
\label{fig:two}
\end{minipage}
\begin{minipage}{0.5\hsize}
\begin{center}
\includegraphics[width=80mm]{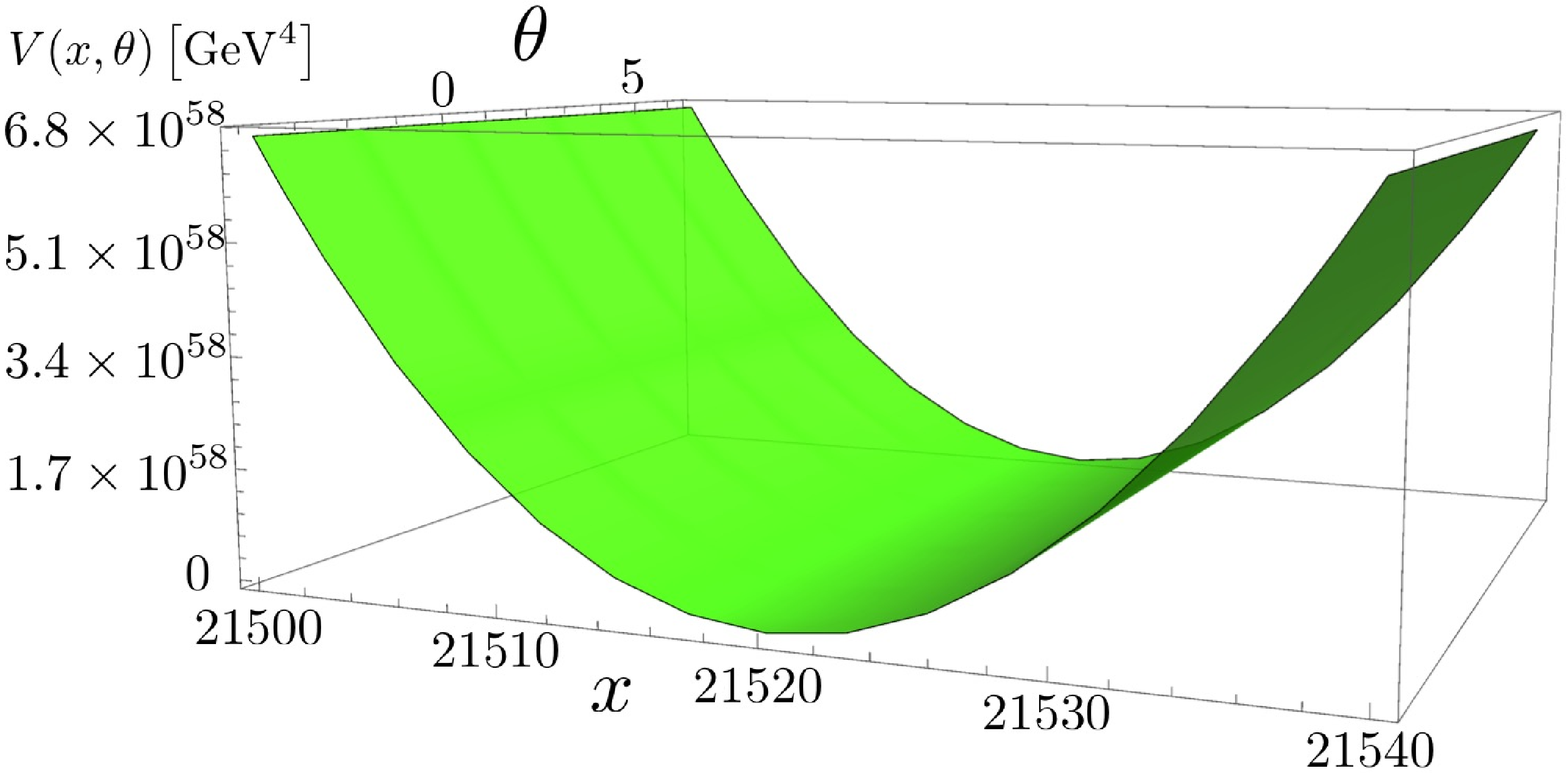}
\end{center}
\vspace{-5mm}
\vskip-\lastskip
\caption{Potential for $r_m=0.1$}
\label{fig:three}
\end{minipage}
\end{figure}

The value of $r_V$ is sensitive to that of $r_m (= \mu/m)$,
and $r_m$ is restricted as $r_m \lesssim 0.3$ for $r_V \lesssim O(10^{-2})$.
In other words, 
the shape of $V$ changes drastically depending on the value of $r_{m}$.
As a reference, typical configurations of $V(x,\theta)$ are
described in Fig.\ref{fig:two} and \ref{fig:three}.
Note that the direction of $x$-axis is zoomed out.
We take the following values for parameters: $r_m=1.0$, $L=6\times10^{-20}{\rm GeV^{-1}}$ 
and $m=5.1\times10^{18}{\rm GeV}$ in Fig.\ref{fig:two},
and $r_m=0.1$, $L=6\times10^{-20}{\rm GeV^{-1}}$ 
and $m=5.1\times10^{18}{\rm GeV}$ in Fig.\ref{fig:three}.
\vspace{0mm}
As seen from these figures,
the values of both $V$ and $x(\theta)$ satisfying $\partial V/\partial x = 0$
change rapidly depending on $\theta$ for $r_m = 1.0$
and the values of  $V$ and $x(\theta)$ are almost independent of $\theta$ for $r_m = 0.1$.

\subsection{\it Gauge-Higgs dominated inflation}

We investigate whether the gauge-Higgs dominated inflation 
(an extension of the extranatural inflation) 
is realized based on $V(x, \theta)$.
We study a case with $c_{1}=4$, $c_{2}=1$ and $r_m = 0.3$.
The reason why we choose $r_m = 0.3$ is 
that the magnitude of $L^{-1}_{\rm phys}$, $m_{\rm phys}$ and $\mu_{\rm phys}$ 
can be larger than the reduced Planck mass if $r_m$ is smaller than $0.3$.
Although the leading term ($k=1$) in $V(x, \theta)$ is known 
to be a good approximation to $V$ itself,
we carry out the numerical analysis including the next-to-leading term ($k=2$)
because it contributes dominantly to 
the second derivative of the potential with respect to $\varphi$ 
(the slow-roll parameter $\eta$) around $\theta=\pi/2$ \footnote{If one considers the running of spectral index that corresponds to third and higher derivatives of the potential with respect to the inflaton,
the higher-order terms ($k>2$) are necessary to evaluate it correctly~\cite{KCC}.}. 

\begin{figure}[h!]
\begin{center}
\includegraphics[width=100mm]{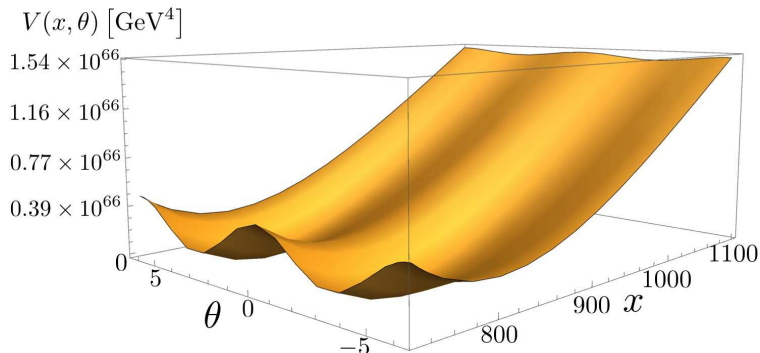}
\vspace{-5mm}
\vskip-\lastskip
\caption{The potential $V(x,\theta)$ for 
$L=6\times10^{-20}{\rm GeV^{-1}}$ and $m=5.1\times10^{18}{\rm GeV}$.}
\label{fig:one}
\end{center}
\end{figure}

The minimum of the potential is given by the conditions,
\begin{eqnarray}
\left.V(x, \theta)\right|_{x=\langle x \rangle,\theta=\pi}=0~,~~
\left.\frac{\partial}{\partial x}V(x, \theta)\right|_{x=\langle x \rangle,\theta=\pi}=0~,
\label{minimum conditions}
\end{eqnarray}
where we use the fact that $\theta$ is fixed as $\theta=\pi$ from $\partial V/\partial\theta=0$.
From (\ref{minimum conditions}),
$a$ and $\langle x \rangle$ are determined as
$a\simeq2.2\times10^{-6}m^{5}$
and $\langle x \rangle \simeq 788$.\footnote{We drop the infinite part of $a$ and show only the finite part. We need a fine tuning to determine $a$, and this is a sort of fine-tuning problem known as the cosmological constant problem.}
As an illustration, the $V(x,\theta)$ is depicted in Fig.\ref{fig:one},
for a specific value of $L$ and $m$. 
\vspace{0mm}

From the conditions (\ref{slow-roll para}) -- (\ref{Power spectrum}),
we obtain the allowed region of $f$ as
\begin{eqnarray}
9M_{G}\lesssim f\lesssim15M_{G}~.
\label{f}
\end{eqnarray}
Hereafter, we use $f=10M_{G}$ as a typical value.
Then, the initial value of $\theta$ is determined as $\theta_{\ast}=1.7$ from (\ref{spectral index}),
and the value at the end of slow-roll is fixed as $\theta_{e}=3.0$ with $N\simeq58$,
using $\epsilon\sim1$ and $\eta\sim1$.
In terms of $\varphi$, we have $\varphi_{\ast}=4.1\times10^{19}{\rm GeV}$ and $\varphi_{e}=7.2\times10^{19}{\rm GeV}$ from $\varphi=f\theta$.

The inflaton mass $M_{\varphi}$ is given by
\begin{eqnarray}
M_{\varphi}^2 =\left. \frac{\partial^{2} V}{\partial\varphi^{2}}\right|_{x = \langle x \rangle,
\varphi = \langle \varphi \rangle}
\label{inflatonmass-formula}
\end{eqnarray}
and its magnitude is estimated as
\begin{eqnarray}
M_{\varphi}=\left(3.2\times10^{-48}L^{2}m^{6}{\rm GeV^{2}}\right)^{1/2}~,
\label{inflatonmass}
\end{eqnarray}
where we use (\ref{deriv2}) with $c_2 = 1$,
$\langle x \rangle \simeq 788$, 
$\langle \varphi \rangle \simeq 7.2\times10^{19}{\rm GeV}$
($\langle \theta \rangle \simeq 3.0$)
and $f = 10 M_G$.
From (\ref{Power spectrum}), $L^{2}m^{6}$ is estimated as
\begin{eqnarray}
L^{2}m^{6}\simeq6.6\times10^{73}{\rm GeV^{4}}~.
\label{Lm}
\end{eqnarray}
Using (\ref{inflatonmass}) and (\ref{Lm}), 
the value of $M_{\varphi}$ is fixed as
$M_{\varphi} \simeq1.5\times10^{13}{\rm GeV}$.
Furthermore, we can estimate the VEV of $\chi$,
using the relation $x =L^{3}m^{3} \chi$
and $\langle x \rangle \simeq 788$, as
\begin{eqnarray}
\langle\chi\rangle\simeq 788 L^{-3}m^{-3}~.
\label{Lm<??>}
\end{eqnarray}
From (\ref{Lm}) and (\ref{Lm<??>}),  the physical mass of $U(1)$ charged fermions 
is determined as 
$m_{\rm phys} = m \langle \chi \rangle^{-1/6} \simeq9.3\times10^{17}{\rm GeV}$.
As a reference, we estimate the masses of other particles.
The radion mass $M_{\chi'}$ is given by
\begin{eqnarray}
M_{\chi'}^2 =\left. \frac{\partial^{2} V}{\partial\chi'^{2}}\right|_{x = \langle x \rangle,
\varphi = \langle \varphi \rangle}
\label{radionmass-formula}
\end{eqnarray}
and its value is fixed as
$M_{\chi'} \simeq 6.7\times10^{15}{\rm GeV}$.
The masses of Kaluza-Klein (KK) modes are roughly given by
\begin{eqnarray}
m^2_{\rm KK} \sim \frac{4 \pi^2 n^2}{\langle\chi\rangle L^2}
= \frac{4 \pi^2 n^2}{\langle\chi\rangle^{1/3}L_{\rm phys}^2}~,
\label{KKmass}
\end{eqnarray}
where $n$ is a positive integer.
The mass of  the first KK mode ($n=1$) is evaluated as
$m_{\rm KK}|_{n=1} \simeq 6.4\times10^{17}{\rm GeV}$.

Next, we impose the following conditions on parameters
in order to limit the range of $L$.
\begin{enumerate}
\renewcommand{\labelenumi}{\roman{enumi})}
\item The value of 4D gauge coupling constant $g_{4}$ is less than unity,
in order to make the analysis based on perturbation trustworthy.
\item The physical fermion masses $m_{\rm phys}$ and $\mu_{\rm phys}$ are 
smaller than the 5D reduced Planck mass $M_{G_5}$ defined by
\begin{eqnarray}
M_{G_{5}}\equiv \sqrt[3]{\frac{1}{8\pi G_{5}}}= \sqrt[3]{\frac{1}{8\pi LG}}~,
\label{5D Planck mass}
\end{eqnarray}
where  $G_{5}$ is the 5D Newton constant.
From the perspective of 5D theory, $M_{G_{5}}$ would be 
more fundamental than the 4D one.
\end{enumerate}
From i) and ii), we obtain the upper and lower bounds on $L$,
\begin{eqnarray}
4.2\times10^{-20}{\rm GeV^{-1}}\lesssim L\lesssim6\times10^{-18}{\rm GeV^{-1}}~,
\label{Lrange}
\end{eqnarray}
where we use $f = 1/(g_4 L) = 10 M_G$
and $m_{\rm phys} \simeq9.3\times10^{17}{\rm GeV}$.
From, (\ref{Lm}), (\ref{Lm<??>}) and (\ref{Lrange}), we have
the upper and lower bounds on $m$ and $\langle\chi\rangle$,
\begin{eqnarray}
1.1\times10^{18}{\rm GeV}\lesssim \hspace{-0.3cm}&m&\hspace{-0.3cm}
\lesssim 5.8\times10^{18}{\rm GeV}~,\label{mrange}\\
2.7\lesssim\hspace{-0.3cm}&\langle\chi\rangle&\hspace{-0.3cm}
\lesssim 55000~.\label{chi-range}
\end{eqnarray}
Note that one of $L$, $m$ and $\langle\chi\rangle$ is a free parameter.
The values of physical parameters are summarized in Table \ref{table:one}.
\begin{table}[h!]
\begin{center}
\begin{tabular}{| c |}
\hline
Values of physical parameters\\ \hline
$L_{\rm phys}^{-1} =1.2 \times10^{17}\sim 6.2 \times10^{17}{\rm GeV}$  \\
$m_{\rm phys}\simeq9.3\times10^{17}{\rm GeV}$ \\
$\langle\chi'\rangle=9.7 \times10^{17}\sim 1.1\times10^{19}{\rm GeV}$\\
$\mu_{\rm phys} \simeq 2.8\times10^{17}{\rm GeV}$ \\
$M_{\varphi} \simeq 1.5\times10^{13}{\rm GeV}$ \\
$M_{\chi'} \simeq 6.7\times10^{15}{\rm GeV}$ \\
$g_{4}= 0.007 \sim 0.9$ \\
$M_{G_{5}}=9.9\times10^{17}\sim 4.9\times10^{18}{\rm GeV}$ \\
$r \simeq 0.10$\\ \hline
\end{tabular}
\caption{The allowed values of physical parameters for $f=10M_G$ and $r_m=\mu/m=0.3~$.}
\label{table:one}
\end{center}
\end{table}
We see that $L^{-1}_{\rm phys}$, $m_{\rm phys}$ and $\mu_{\rm phys}$ are obtained 
as sub-Planckian quantities depending on the value of radion, 
even if the original parameters $L^{-1}$, $m$, $\mu$ are trans-Planckian.
The $g_{4}$ can be as large as $0.9$
even if $f$ is considerably larger than $M_G$.
It is in contrast to the result in the original extranatural inflation, i.e., 
$g_{4}$ turns out to be tiny of $O(10^{-2})$~\cite{ACCR}.
The reason why $g_{4}$ can be unity in our model 
is that $\langle\chi\rangle$
can be much larger than $O(1)$, as seen from the relation
\begin{eqnarray}
g_4 = \frac{1}{f L} = \frac{\langle\chi\rangle^{1/3}}{f L_{\rm phys}}~.
\label{g4-L}
\end{eqnarray}

The tensor-to-scalar ratio \eqref{tensor} is evaluated as $r \simeq 0.10$. 
It is within the range of observational upper bound, $r<0.12$~\cite{Planck2015},
but still large enough to be testable by the future observation.

From the above analysis,
it is confirmed that the gauge-Higgs has properties required for the inflaton 
and the gauge-Higgs dominated inflation can be achieved not so unnaturally.

\section{\it Conclusion and discussion}

Using the radion gauge-Higgs potential obtained from one-loop corrections
in the 5D gravity theory 
coupled to a $U(1)$ gauge boson and matter fermions,
we have found that the gauge-Higgs $\varphi$ can give rise to a large-field inflation
in accord with the astrophysical data. 
In contrast, it is difficult to realize the inflation dominated by the radion $\chi'$,
because the slow-roll conditions cannot be fulfilled except for a narrow region satisfying
$\partial V/\partial \chi' \simeq 0$.
Furthermore, the hybrid inflation scenario is not achieved in our model
that $\chi'$ is moving toward the minimum of the potential in the first stage.

Based on the gauge-Higgs dominated inflation scenario,
we have determined the values of parameters 
using constraints on the inflaton potential. 
Some of them are different from those 
in specific gauge-Higgs inflation models~\cite{ACCR, IKLM}, 
and this is mainly due to the difference of setup.
Our model contains the gravity
and the physical parameters such as the 4D gauge coupling constant $g_4$,
the size of extra space $L_{\rm phys}$, 
fermion masses $m_{\rm phys}$ and $\mu_{\rm phys}$ are 
multiplied by some power of $\langle \chi \rangle$.
Even if the parameter $f$
is trans-Planckian, i.e., $f > M_G$, $g_4$ can be as large as 
the standard model gauge coupling constants at the grand unified scale.
This is due to the fact that $g_4$ is
proportional to $\langle \chi \rangle^{1/3}$
and $\chi$ can take a pretty large VEV.
In this case, the size of $L_{\rm phys}^{-1}$, $m_{\rm phys}$ and $\mu_{\rm phys}$ 
can be below the 5D reduced Planck mass.
As seen from the fact that $\varphi$ works as the inflaton,
$\varphi$ acquires the mass of $O(10^{13})$GeV through radiative corrections.
Other massive particles are much heavier than $\varphi$.
The masses of the first KK modes and the fermion zero modes
are of $O(10^{17})$GeV
and the radion mass is of $O(10^{15})$GeV.
Because the value of radion stays almost constant
in the region with $\partial V/\partial \chi' \simeq 0$,
the physical size of extra dimension has been almost stabilized
at the initial time of inflation
and $V(\langle \chi' \rangle, \varphi)$ is treated
as the inflaton potential effectively.

Our model is left problems concerning parameters.
The first one is a common problem in inflation models,
how the inflaton can take a suitable initial value
to realize the inflation compatible with the observation.
Because our potential has $\chi'$ and $\varphi$,
it can be traded for the initial-value-problem of $\chi'$
that is our future work.
The second one is to determine the value of $f$ without using
constraints on the inflation potential.
It is necessary to fix $g_4$ and $\langle \chi \rangle$.
The value of $g_4$ can be obtained by the VEV of some scalar field
such as the dilaton and/or the moduli.
It would be efficient to extend our system by incorporating such scalar fields.
The magnitude of $\langle \chi \rangle$ strongly depends on
the ratio of fermion masses $r_m = \mu/m$.
For the smaller $r_m$, the larger $\langle \chi \rangle$ is obtained.
It would be important to explore the origin of massive fermions.
The last one is how our results are reliable after receiving
gravitational corrections, which are uncontrollable at present.

Furthermore, the effective field theory at the Planck scale has not yet been known.
In our model, the values of several parameters can be 
of the order of the Planck mass and the field value of the gauge-Higgs is above the Planck scale.
This result could indicate that the quantum theory of gravity such as string theory
is necessary to understand the mechanism of inflation more properly.
Using it, there is a possibility that gravitational corrections are controlled and our analyses are justified.
In string theory construction of inflation models, the type of axion inflation is extensively studied (e.g., see \cite{MSWW}) for large-field inflation. We note that our effective potential is generated through the perturbative loop corrections and the origin of inflaton(s) is different from that in axion inflation models, though some of the properties of the inflaton potential, especially the periodicity, are shared.
It would be interesting to study the inflation
based on the effective potential relating several scalar fields
such as the dilaton, the moduli (including the radion) and the gauge-Higgs
in the framework of string theory.

\section*{Acknowledgement}
~~~~This work is supported in part by funding from Nagano Society for The Promotion of Science (Y.A.),
scientific grants from the Ministry of Education, Culture, Sports, Science and Technology 
under Grant Nos. 21244036 and 20012487 (T.I.), 
Grant Nos. 21244036 and 22540272 (Y. Kawamura), 
and the National Center for Theoretical Science (NCTS) 
and the grant 101-2112-M-007-021-MY3 of 
the Ministry of Science and Technology of Taiwan (Y. Koyama). 
Y.A. benefited greatly from his visit to National Taiwan University. 
He wishes to thank Professor Ho for giving him the opportunity to visit. 
T.I. wishes to thank CTS of NTU for partial support.


\end{document}